# IDENTIFYING DARK MATTER THROUGH THE CONSTRAINTS IMPOSED BY FOURTEEN ASTRONOMICALLY BASED "COSMIC CONSTITUENTS"

Jerome Drexler
NJIT Research Professor
New Jersey Institute of Technology

**Abstract**

Mankind has not yet explained dark matter, the accelerating expansion of the Universe, the "knee" and "ankle" of the cosmic ray energy spectrum graph, the low star formation rates of low surface brightness (LSB) dwarf galaxies, the ignition of hydrogen fusion reactions in the first generation stars or the departing locations of earthbound high-energy cosmic ray protons. A new research hypothesis has been developed by the author based upon finding astronomically based "cosmic constituents" of the Universe that may be created or influenced by or have a special relationship with possible dark matter candidates. A list of 14 relevant and plausible "cosmic constituents" of the Universe was developed by the author, which was then used to establish a list of constraints regarding the nature and characteristics of the long-sought dark matter particles. A dark matter candidate was then found that best conformed to the 14 constraints established by the "cosmic constituents." The author then shows that this same dark matter candidate provides plausible explanations for the accelerating expansion of the Universe, dark energy, both the "knee" and "ankle" of the cosmic ray energy spectrum graph, the low star formation rates of LSB dwarf galaxies, the ignition of hydrogen fusion reactions in the first generation stars, the source of magnetic fields in spiral galaxies, how dust particles could facilitate hydrogen fusion in stars and the four departing locations, within the Local Group and Virgo Supercluster, for earthbound high-energy cosmic ray protons.

**Introduction**

Some cold dark matter (CDM) researchers believe that considerable progress has been made during the past 20 years [1]. They believe that the cold dark matter theory fits the cosmic microwave background data, explains why the observed galaxies have specific sizes and numbers and has been used successfully in predicting galaxy distributions in the Universe. They also believe that the CDM theory makes correct predictions about galaxy growth, mergers, shapes of dark matter halos and galaxy clusters and feel that for any new competing theory of dark matter and structure formation to be considered for adoption, it would have to compete successfully against the CDM theory.

In order to evaluate a competing new dark matter theory in a reasonable amount of time, the author proposes that the initial research effort should lean heavily on analysis of existing astronomical data and on multi-faceted comparisons with the weakly interacting massive particles (WIMPs) and neutralinos of the CDM theory. The author believes that a combined scientific and logical approach should be employed that utilizes a group of astronomically based "cosmic constituents" and the constraints they place on the nature and characteristics of dark matter to identify dark matter and to highlight the strengths and weaknesses of the competing dark matter candidates.

The proposed initial research direction is somewhat based upon the logical system called Ockham's Razor, which was developed by the English logician, William of Ockham, in the 14$^{th}$ Century. Simply described, Ockham's Razor logic states that a researcher should not make more assumptions than absolutely necessary in seeking explanations for observed phenomena. That is, if there are two theories proposed to explain a scientific phenomenon, the theory that requires the fewer assumptions to explain it should be favored. If there are two competing theories and only one astrophysical or cosmological mystery, it could be difficult to select the winning theory on the basis of that one mystery alone. In this case, additional astronomical data or scientific mysteries should be employed to facilitate the competition.

If one were to assemble a list of astronomically based astrophysical and cosmological phenomena that have relevance to the competing dark matter theories, we might be able to subject the theories to a pentathlon or decathlon competition in order to crown the winning theory. To utilize this approach, the author has selected a group of 14 "cosmic constituents" along with their inherent constraints on the nature and characteristics of dark matter for a multi-faceted analysis now and for a competition with other dark matter candidates in the future. None of these 14 "cosmic constituents" involves any aspects of the early Universe, which would of necessity involve *unproven theories that are equivalent to assumptions*.

A "cosmic constituent" is an astronomically based cosmic phenomenon such as dark matter halos, the accelerating expansion of the Universe, the rotation curves of galaxies or ultra-high-energy cosmic rays that might be influenced by a Universe filled 80% to 90% with some type of dark matter. Each of these "cosmic constituents" constrains the characteristics or nature of dark matter such as the way that Fritz Zwicky *[2]* did with the rotation curves of galaxies within galaxy clusters and Vera Rubin *[3] [4]* did with rotation curves of stars and atomic hydrogen within galaxies. That is, if a group of "cosmic constituents" and the associated group of constraints is chosen to try to identify dark matter, then the "signature characteristics" of any considered dark matter candidate would have to be compatible with this same group of constraints.

The first goal was to select the relevant astronomically based "cosmic constituents" that may be created by, influenced by or have a special relationship with any potential dark matter candidates. The second goal was to utilize the "cosmic constituents" and the astrophysical constraints they impose on any dark matter candidates to screen out non-conforming candidates and to select the most logical choice. The third goal was to provide a descriptive set of characteristics of the selected dark matter candidate in the form of a group of "signature characteristics." Finally, the "signature characteristics," which totaled 24, of the selected dark matter candidate were utilized in providing a plausible explanation of the nature of each of the "cosmic constituents" and/or its relationships with the selected dark matter candidate. It turned out that this overall approach may have yielded some possible astrophysical discoveries.

**Additional Approaches to Dark Matter Research**

Previously, dark matter research focused entirely on uncharged particles. The research approaches that had been used included:

1. Conducting computer simulations based upon uncharged particles.

2. Determining the ratios of dark matter to ordinary matter in galaxies and clusters.

3. Exploring the reasons for the flat rotation curves of spiral galaxies.

4. Utilizing special detectors in searching for uncharged WIMPs and neutralinos.

5. Employing gravitational lensing utilizing intervening dark matter distributions.

6. Measuring galaxy cluster motion to reveal presence of dark matter superstructure.

7. Recording and analyzing the mass density ripples in the early Universe via the cosmic microwave background (CMB).

Since dark matter is responsible for about 80% to 90% of the mass of the Universe, it is likely to influence or have relationships with a number of "cosmic constituents" that have not been adequately utilized or explored in prior research. The development of additional research approaches to uncover the nature of dark matter was guided by the following three related hypotheses:

1. The Universe comprises "cosmic constituents" formed into a system arrangement in which some of the principal "cosmic constituents" are mutually dependent.

2. Some of the principal mysteries of astrophysical cosmology are linked, making it possible to find a single unique astrophysical theory to solve several of the cosmological mysteries.

3. Since traditional cosmology and astrophysics have failed in identifying dark matter, the source of high and ultra-high energy (UHE) cosmic rays, dark energy and the accelerating expansion of the cosmos and how galaxies are formed, perhaps an expanded approach to dark matter research should be tried. That is, in researching the nature of dark matter, an analysis of its influence on and its relationships with 14 relevant "cosmic constituents" of the Universe might provide important clues as to some of the characteristics of dark matter.

**Dark Matter Research Guided by the Three Related Hypotheses**

The dark matter research approaches utilized by the author were guided by the previously mentioned three related hypotheses which included the selection of the "cosmic constituents" of the Universe that possibly could be strongly influenced by or have a relationship with the enormous volume of dark matter particles in the Universe.

The potentially relevant "cosmic constituents" that were hypothesized to be influenced by the multitudinous dark matter particles also included dark energy, new star ignition, the spherical dark matter halos around spiral galaxies, the extragalactic magnetic fields, the flat rotation curves of spiral galaxies, the linearly rising rotation curves of LSB dwarf galaxies and the ratio of dark matter to ordinary matter in the Universe.

The next step after selecting a group of relevant "cosmic constituents" was to try to identify the type of dark matter particles that could create, cause, form, lead to, influence or have a relationship with many or most of the selected "cosmic constituents." An evolving series of rhetorical questions was considered along the way, leading to the following list:

*What type of dark matter particles could:*

1. Form spherical dark matter (DM) halos around galaxies and DM halos around galaxy clusters?

2. Cause the accelerating expansion of the Universe and possibly store dark energy?

3. Be transformed into low velocity hydrogen, protons or proton cosmic rays?

4. Create the magnetic fields within and around spiral galaxies?

5. Be concentrated in the long large curved filaments of dark matter (announced by NASA on September 8, 2004), which form galaxy clusters where two DM filaments intersect?

6. Create large mature spiral galaxies less than 2.5 billion years after the Big Bang?

7. Create spherical DM halos having predictable outer and "hollow" core diameters?

8. Provide angular momentum to spiral galaxies and DM halos?

9. Create galaxies without a central DM density cusp?

10. Create a starless galaxy or a LSB dwarf galaxy with low star formation rates?

11. Lead to linearly rising rotation curves for LSB dwarf galaxies and to flat rotation curves for spiral galaxies?

12. Form about 80% to 90% of the mass of the Universe, the remainder being hydrogen, helium, etc.?

13. Ignite hydrogen fusion reactions of second generation stars utilizing hydrogen molecules and dust and ignite fusion reactions of the first generation stars with only hydrogen atoms?

14. Create the first "knee" at $3 \times 10^{15}$ eV, the second "knee" between $10^{17}$ eV and $10^{18}$ eV and the ankle at $3 \times 10^{18}$ eV of the cosmic-ray energy spectrum near the Earth?

It appears that there is at least one dark matter candidate that could have an influence on or relationship with at least ten of the above-indicated "cosmic constituents"—namely, relativistic protons, accompanied by helium nuclei, that are constrained by the galactic and extragalactic magnetic fields into Larmor Radius spiral orbits forming dark matter halos around galaxies and around galaxy clusters and that also are concentrated in long large curved filaments of dark matter.

**Orbiting Relativistic Protons, Accompanied by Helium Nuclei, Hypothesized to Be the Long-sought Dark Matter Particles, Are Expected to Have the Following Signature Characteristics (SigChar), Denoted as SigChar A through X:**

A. *DM proton energies.* Relativistic protons in DM halos would have energies ranging from about $10^{15}$ eV to $10^{20}$ eV. *[5]* Their energies would decline continuously through synchrotron radiation losses leading to a proton flow into the associated galaxy. In the case of the Milky Way the galactic magnetic field is 2,000 times greater than the extragalactic magnetic field. Therefore protons entering it would experience soaring synchrotron radiation energy losses and plunging kinetic energies with the result that only a portion of the protons entering a star system (as cosmic ray protons) would still be relativistic. Note that the relativistic protons in a galaxy halo are the dark matter, and their flux would be orders of magnitude greater than the flux of the cosmic ray protons entering all the star systems of the same galaxy. Note, also, that although there is a strong limit from primordial nucleosynthesis on the maximum baryonic *particle density* in the Universe, the high relativistic mass of the dark matter protons provides the necessary DM *mass density* while staying within the baryonic particle density limits.

B. *Milky Way's magnetic fields.* The DM protons move through the extragalactic magnetic field of about $10^{(-9)}$ gauss in the dark matter halo of the Milky Way and through the 2,000 times higher galactic magnetic field of about $2 \times 10^{(-6)}$ gauss. *[6]*

C. *Larmor Radius equation.* The DM protons' spiral paths in the DM halo are determined by the Larmor Radius equation *[6]*, as follows:

$$r = 110 \text{ Kpc} \times \frac{10^{-8} \text{ gauss}}{B} \times \frac{E}{10^{18} \text{ eV}}$$

where Kpc means kilo parsec and
one parsec equals 3.26 light-years

D. *Milky Way's DM halos and proton energies.* The above three SigChars plus the diameters of the Milky Way, its dark matter halo and its Local Group's halo could lead to an estimate that relativistic protons in the Milky Way's dark matter halo would have energies of about $3 \times 10^{15}$ eV at its inner core diameter and about $3 \times 10^{17}$ eV at its outer diameter. The dark matter halo of its Local Group galaxy cluster probably would have protons with energies of above $3 \times 10^{17}$ eV at its inner core diameter and about $6 \times 10^{18}$ eV at its outer diameter. The DM halos in and around the Virgo Supercluster probably would contain protons with energies above $6 \times 10^{18}$ eV. In the well-known "leaky box" model for cosmic rays, higher energy protons escape from a galaxy into its halo. In contrast, in this model of a hierarchy of DM halos enclosing smaller DM halos, a "leaky halo" model is employed wherein relativistic protons that have lost kinetic energy for various reasons leak into the next smaller lower energy halos and finally into a star system of a galaxy as *cosmic rays*. Therefore, the cosmic ray protons arriving at the solar system would tend to be the lowest energy protons in the Milky Way and in its DM halo and in the DM halos of the Local Group and the Virgo Supercluster. Also, the higher the energy of the arriving cosmic rays the greater the distance they had traveled to reach the solar system. Thus, the fall off of proton flux with proton energy of the arriving cosmic rays could be much more rapid than that for the DM protons residing in the three halos and the galaxy. See item (14) of Tentative Conclusions for the diameters of the Milky Way, its halo, etc. *[5] [7] [8]*

E. *Paths of protons.* The orbits and paths of relativistic protons would be determined by their kinetic energies, the local magnetic field strengths and the Larmor Radius equation. Their gravitational accelerations from nearby mass objects should be relatively negligible since gravitational accelerations are many orders of magnitude smaller than electro-magnetic accelerations. *[6]* Synchrotron radiation losses would continuously reduce the radii of the proton paths, causing DM protons to move from a large diameter DM halo into its enclosed galaxy and eventually into star systems as cosmic ray protons.

F. *Proton streams creating magnetic fields.* The dark matter halos surrounding spiral galaxies, even very young galaxies, exhibit magnetic fields. The relativistic protons of the author's Dark Matter Theory follow spiral paths that are determined by a halo's magnetic field strength, and these orbiting protons also participate in creating and maintaining the halo's magnetic field strength via the astrophysical dynamo effect, as described in "The Magnetic Universe─Geophysical and Astrophysical Dynamo Theory." *[ 9]* Also see A&A article entitled, "Strong magnetic fields and cosmic rays in very young galaxies." *[10] [11]* Other researchers reported on March 9, 2005, "There are several viable alternatives to explain the coherent magnetic fields that we observe [in the LMC]. Potentially most pertinent for the LMC (Large Magellanic Cloud) is the cosmic-ray driven dynamo …." *[12] [13]*

G. *Proton flux and kinetic energy in halos.* An analysis of SigChars A through F (particularly SigChar D) and the knowledge that cosmic-ray proton flux density in the Universe declines rapidly with proton kinetic energy, leads to the conclusion that a galaxy's dark matter halo would be expected to have the highest proton flux density and lowest kinetic energy protons near its inner core (near the galactic magnetic field) and the lowest proton flux density and highest energy protons near its outer diameter (in the extragalactic magnetic field). *[5] [6]*

H. *Proton relativistic mass losses from synchrotron radiation.* Whenever relativistic protons move across a magnetic field, they are deflected (accelerated) and radiate photons as synchrotron radiation. The smaller the radius of curvature of the spiral path of a proton, the more photon energy it emits as synchrotron radiation, which causes the proton's kinetic energy, velocity and relativistic mass to decline. Such orbiting DM halo protons with declining relativistic mass would cause the gravitational acceleration on nearby galaxies or gas clouds to decline. *[14]* (Note that a proton's kinetic energy losses from synchrotron radiation are lower by a factor of about 11 trillion compared to that of an electron. Therefore, some DM protons can remain in DM halos for extremely long periods of time provided they do not collide with photons, dust or hydrogen or helium.)

I. *Magnetic bulges leading to increased synchrotron radiation from protons.* Whenever a magnetically constrained relativistic proton encounters a rise in the magnetic field, its acceleration increases and its synchrotron radiation energy losses increase and as a result its velocity, relativistic mass and radius of curvature of its spiral path would tend to decline. *[14] [15]*

J. *Why DM halo protons enter its galaxy.* It follows from the previous paragraph that a stream of magnetically constrained relativistic protons in a dark matter halo skimming the surface of its galaxy would experience the same effects owing to the higher magnetic field of the galaxy, thereby causing the proton stream to move deeper into the galaxy and leading to increased synchrotron radiation and a loss of kinetic energy and relativistic mass. By this means, the combined mass of the galaxy and its DM halo would decline. (A similar process can occur if the halo protons collide with dust, photons or hydrogen/helium clouds causing the protons to slow down, thereby decreasing their path radius and increasing their synchrotron radiation energy losses.) *[14] [15]*

K.  *Proton/helium nuclei collisions with hydrogen clouds*. High velocity collisions by the relativistic protons (and by helium nuclei) with atomic hydrogen in compressed interstellar hydrogen clouds in a galaxy could lead to a hydrogen fusion reaction and the ignition of new stars. Note that colliding protons with energies of $10^{15}$ eV would be 1000 times more powerful than those produced by today's man-made accelerators and therefore could play a more significant role in initiating and maintaining hydrogen fusion reactions. *[16]*

L.  *Linearly rising rotation curves indicating that LSB dwarf galaxy DM halos are "weakly centrally concentrated" (that is, "hollow")*. A paper on LSB halos, authored by J. Bailin et al and published in MNRAS 14 February 2005 *[17]*, contains some pertinent statements: "These studies focused on the shape of the LSB galaxy rotation curves, which rise approximately linearly with radius in a manner one expects if the mass density profile contained only a shallow central cusp or a constant density core. … Thus, a slowly rising rotation curve could be interpreted as indicative of a halo that is weakly centrally concentrated."  An excerpt from the abstract reads, "We find that our LSB galaxy analogues occupy haloes that have lower [mass] concentrations than might be expected …."  Below, SigChar T provides a theory for the LSB dwarf galaxies' low star production rates and in the process translates the expression "weakly centrally concentrated" LSB DM halo into the simpler term "hollow" LSB DM halo.  *Flat rotation curves of spiral galaxies.* The well-known flat rotation curves of stars and hydrogen in spiral galaxies indicate that the spherical mass of dark matter, contained within a sphere of radius $r$ from the nucleus of the spiral galaxy, increases linearly with radius $r$ through the galaxy and into the halo. *[3] [4]*  This means that the dark matter mass density at radius $r$ must be declining approximately as the square of the radius $r$. Relativistic proton streams in dark matter halos, as pointed out in SigChar G above, are expected to have a rapidly declining mass density with radius, that is, the highest proton mass flux density would be near the hollow inner core of the DM halo and the lowest proton mass flux density would be near the DM halo's outer diameter. More specifically, while the *particle* flux density of the relativistic protons falls as a power law of their energies with the exponential decline ranging between 2.7 and 3 as determined from the cosmic-ray energy spectrum graph *[5] [18]*, the relativistic mass of each proton rises linearly with kinetic energy and is highest at the DM halo's outer diameter. *[5]*  This leads to the approximation that the *mass* flux density, within the galaxy and its halo, falls nearly inversely as the square of the radius $r$ from the nucleus of a spiral galaxy, thereby approximately satisfying Vera Rubin's flat-rotation-curves requirement and linking this achievement with the distribution of proton mass and kinetic energy distribution within the DM halo.

M.  An e*xplanation for the two "knees" and "ankle" of the cosmic ray energy spectrum*.  The relativistic proton dark matter theory may explain the two "knees" of the cosmic-ray energy spectrum graph at $3x10^{15}$ eV and at about $3x10^{17}$ eV and at the "ankle" of the graph at $3x10^{18}$ eV in terms of four different cosmic-ray source locations. See SigChar D and *[5] [7] [8] [18]*.  The departing location of most of the cosmic-ray protons approaching the Earth with energies between $10^9$ eV and $3x10^{15}$ eV is probably the Milky Way since the relatively strong galactic magnetic field confines these lower energy cosmic rays to the Galaxy. Most of the Earth-arriving cosmic-ray protons with kinetic energies between $3x10^{15}$ eV and about $3x10^{17}$ eV probably come via the Milky Way's DM halo through the Milky Way to the solar system for reasons explained in SigChar J. The galactic magnetic field is too weak to prevent them from reaching the solar system. Most of the cosmic-ray protons with energies between $3\times10^{17}$ eV and $6\times10^{18}$ eV probably arrive at the Earth via the DM halo surrounding our Local Group galaxy cluster for reasons similar to those in SigChar J and because the intervening magnetic fields are too weak to prevent these ultra-high energy relativistic protons from passing through the galaxy cluster, the Milky Way's DM halo and the Milky Way before finally reaching the solar system. The DM halos in and around the Virgo Supercluster probably would contain protons with energies above $6\times10^{18}$ eV and probably would be the departing location of the highest energy cosmic rays arriving at the Earth.. This latter conclusion appears to be supported by the fact that at $10^{19}$ eV only 3 to 4 protons per square kilometer arrive at the Earth each century.

N. *Proton synchrotron radiation losses and proton collision losses possibly could lead to the accelerating expansion of the Universe.* The relativistic proton dark matter theory leads to an explanation of the accelerating expansion of the Universe without any additional assumptions. *[14][19]* The relativistic proton DM theory posits that the observed dark matter halos around spiral galaxies and their clusters may be comprised of relativistic-velocity/relativistic-mass protons following halo-size spiral paths through the DM halos as they continually lose relativistic mass due to kinetic energy losses from synchrotron radiation, and collisions with photons, dust and atoms of hydrogen and helium. The well-known galactic and extragalactic magnetic fields would establish both the halo-size Larmor Radius spiral paths of the high-relativistic-mass protons and their relativistic mass losses due to the synchrotron radiation losses. With galaxy clusters already experiencing separation velocities according to the Hubble Law, the decreasing relativistic masses of the DM halos around galaxy clusters and around their galaxies should cause the galaxy cluster separation velocities to increase under the Law of Conservation of Linear Momentum, thereby leading to the accelerating expansion of the Universe. The energy to accomplish this separation acceleration could be derived directly from the conversion to kinetic energy of some of the relativistic mass of the DM halos of the galaxy clusters and the DM halos of the galaxies. In this case, the long-sought mysterious *dark energy* might be related to the total kinetic energy and/or the energy equivalent of the relativistic mass of all the dark matter protons in the various galaxies, galaxy clusters and DM halos.

O. *Radiating DM halo protons become cosmic-ray protons.* When relativistic protons in a DM halo give up some kinetic energy through synchrotron radiation or through collisions with dust, photons or atoms of hydrogen or helium, they decelerate into slower moving relativistic protons. When the kinetic energy of some of the relativistic protons in the Milky Way's DM halo is reduced below $3\times10^{15}$ eV, their synchrotron radiation losses increase and their kinetic energy, relativistic mass and radius of curvature of their spiral orbits decline. The relativistic DM protons then move into the galaxy and eventually plunge into a star system as cosmic ray protons. *[14]*

P. *Long, large DM filaments creating galaxy clusters.* The September 8-9, 2004 news releases from NASA/Harvard entitled, "Motions in nearby galaxy cluster reveal presence of hidden superstructure" regarding Chandra x-ray images of the Fornax cluster states: "Astronomers think that most of the matter in the universe is concentrated in long large filaments of dark matter and that galaxy clusters are formed where these filaments intersect." It should noted that such a filamentary dark matter structure could be a slightly curved portion of a DM halo around or within some galaxy supercluster. This relatively new top-down theory of galaxy cluster formation is compatible with the relativistic proton dark matter theory as described in the author's book published in December 2003. *[5]* (Prior to September 8, 2004, the standard theory of cold dark matter galaxy formation was based upon the bottom-up hierarchical model wherein small galaxies form first and then gravitationally move together over time to form larger galaxies and galaxy clusters.)

Q.  *Mature galaxies in a young Universe.*  The recent discovery of the existence of mature galaxies only about 2.5 billion years after the Big Bang *[20] [21] [22]* (and confirmed by the Carnegie Observatories on March 10, 2005) can be explained using the relativistic proton dark matter theory that involves fast protons that slow down over time due to synchrotron radiation losses, but raises questions about the cold dark matter bottom-up theory of galaxy formation which involves only slow-moving particles. The referenced articles in the July 2004 issue of *Nature* are entitled, "A high abundance of massive galaxies 3-6 billion years after the Big Bang" and "Old galaxies in the young Universe." The Carnegie Observatories had announced in a news release on March 10, 2005 that "Astronomers have found distant red galaxies—very massive and old—in the universe when it was only 2.5 billion years post Big Bang."

R.  *Conservation of angular momentum.*  All spiral galaxies exhibit an angular momentum. This is consistent with the relativistic proton dark matter model in which protons move in halo-size spiral orbits around galaxies and through the galactic and extragalactic magnetic fields according to the Larmor Radius equation. This orbiting stream of relativistic protons possesses a considerable amount of angular momentum to transfer to galaxies and DM halos under the Law of Conservation of Angular Momentum.

S.  *No DM cusps in nuclei of galaxies.* The galactic magnetic field of a spiral galaxy would tend to constrain relativistic DM protons from moving toward the nucleus of a spiral galaxy. When these DM protons lose kinetic energy and move into the galaxy as cosmic-ray protons, they will encounter many star systems as collision targets before reaching the galaxy nucleus. Therefore, these protons should not tend to form a dark matter density cusp at the nucleus of a galaxy.

T.  *An explanation for LSB dwarf galaxies' low star production rates.*  See SigChar L and references *[4] [17]*. If the diameter of a dwarf galaxy disc is smaller than the "hollow" core of its dark matter halo, the number of DM halo protons entering the galaxy to ignite or feed stars would be low or very low. This would occur because the DM halo protons would not be subjected to the full magnitude of the galaxy's higher magnetic field, which typically leads to a rapid rise in synchrotron radiation losses and a decline of the protons' kinetic energy followed by their movement into the enclosed galaxy. The author believes that LSB dwarf galaxies and starless dark galaxies probably have this smaller-galaxy-disc/larger-halo-core relationship. (However, it should be noted that LSB dwarf galaxies are known to be isolated in space, which would eliminate dust accreted onto them from external sources and thus would suggest a lower star production rate.) The author also believes that an LSB or starless galaxy could evolve into a star-creating galaxy as the galaxy grows in mass and size fed by slowed protons from the relativistic protons of the DM halo over millions to billions of years until the galaxy disk becomes equal to or larger in diameter than the "hollow" core of its DM halo and the two celestial bodies overlap. The author believes that the Milky Way is of this overlapping disc-halo type because it is estimated that the dark matter mass reaching into the Milky Way approximately equals the ordinary matter mass of the galaxy and it is a star-forming galaxy. (Also see SigChar D.) One example of a well known LSB dwarf galaxy is DDO154, also known as NGC 4789A and UGC8024, which contains a very large amount of atomic hydrogen gas and has a very large ratio of dark matter to ordinary matter, but for some reasons has a very low star formation rate. Another dwarf galaxy of this type with a huge disk of rotating atomic hydrogen gas, UGC 5288, was studied with a radio telescope by Indiana University using the NSF VLA telescope and reported on in a press release dated January 10, 2005. On February 18, 2005, *Astronomy* magazine published an article entitled, "The first dark galaxy?" in which astronomers from Britain (Cardiff University), Australia, France and Italy say, "A [rotating] cloud of gas in the Virgo cluster may be the first dark galaxy ever found. … The mysterious object has one-tenth the Milky Way's mass but consists of hydrogen gas and dark matter—with no detectable stars." Yet its mass-to-blue-light ratio is at least ten times that of the Milky Way. In December 2004 the University of Virginia announced the discovery of a galaxy named Zwicky 18 which existed as a galaxy in an embryonic state for billions of years and "went through a sudden first starburst only about 500 million years ago." It is possible that this galaxy was smaller in diameter than the "hollow" core of its DM halo for billions of years until they finally overlapped after the galaxy grew in size from accretion of protons, helium nuclei, etc., from its DM halo.

U. *The relativistic kinetic energy of all the relativistic protons in the Universe may be part of the theorized "dark energy."* Professor Pierre Sokolsky wrote the following about cosmic rays in the Universe in his 2004 book entitled, "Introduction to Ultrahigh Energy Cosmic Ray Physics" *[23]*: "If the energy density that we observe on earth is similar to what exists in extragalactic space, a significant component of the total energy of the Universe is in cosmic rays. The cosmic ray energy density integrated over all energies turns out to be approximately 1eV/cm3. For comparison, starlight has an energy density of 0.6 eV/cm3 and the energy density of the galactic magnetic field is 0.2eV/cm3. It is clear that cosmic rays form a major constituent of the interstellar medium." Note that the cosmic ray energy density of 1ev/cm3 only represents the cosmic ray protons and helium nuclei raining on the star systems of a galaxy. The relativistic dark matter protons circulating in Larmor orbits around galaxies and clusters should represent a much greater energy density in the form of kinetic energy, which may be all or part of the "dark energy" that drives the accelerating expansion of the galaxy clusters as described in SigChar N. These proton Larmor orbits also create magnetic fields under the astrophysical dynamo effect that have a significant energy density. See references *[9] [10] [11]*.

V. *Linking relativistic dark matter and dark energy.* If the mass/energy of the Universe is comprised of about 4% ordinary matter, 23% dark matter and 73% dark energy, it probably would be logical to expect a link and perhaps even a coupling between dark matter and dark energy, particularly since a large percentage of the dark matter mass in the Universe appears to be relativistic mass. See SigChars N and U for an example of a possible coupling between dark matter and dark energy.

W. *How the first generation stars may have been ignited without dust or molecular hydrogen.* The following excerpts from a November 18, 2004 article written by the European Southern Observatory (ESO) about starburst galaxies, entitled "Stellar Clusters Forming in the Blue Dwarf Galaxy NGC 5253," may provide a link between the relativistic proton dark matter and primordial star formations. (The related scientific article was published shortly afterward on the astro-ph archives. *[24]*) "Star formation begins with the collapse of the densest parts of interstellar clouds, regions that are characterized by comparatively high concentration of molecular gas and dust like the Orion complex and the Galactic Centre region. Since this gas and dust are products of earlier star formation, there must have been an early epoch when they did not yet exist." The next paragraph of the article continues, "But how did the first stars then form? Indeed to describe and explain 'primordial star formation' without molecular gas and dust is a major challenge in modern astrophysics." A solution to this primordial star ignition problem may be that relativistic DM protons and helium nuclei colliding with compressed interstellar clouds of hydrogen and helium atoms could generate millions of muons to act as catalysts and lead to a hydrogen fusion reaction and the ignition of new stars. The Drexler star ignition and hydrogen fusion theory is based upon an estimate that each $10^{15}$ eV cosmic ray proton striking the Earth's atmosphere produces perhaps hundreds to thousands of muons. See SigChar X. (They actually produce pions which rapidly decay into muons which, in turn, decay less rapidly into electrons, etc., in a number of microseconds.) Also, muons are known to catalyze hydrogen fusion reactions by forming muonic molecular ions comprised of a proton plus a deuteron, helium nucleus, or another proton orbited by a muon. *[25] [26]* Star-related hydrogen fusion might have been feasible in the early Universe since relativistic DM protons were probably multitudinous and had energies much more than a thousand times higher than what can be achieved with man-made accelerators today and catalytic muons were being produced in the millions (about hundreds to thousands of muons per colliding proton) and hydrogen and helium atoms were available as collision targets for the DM protons and helium nuclei spiraling into a galaxy. (Note that this process implies the existence of a high velocity muon-electron-proton plasma.) See SigChar J, K and also X, which discusses the role of the muons. Also see SigChar O regarding collisions of the relativistic dark matter protons with dust, photons and hydrogen.

X. *How the later generations of new stars may have been ignited utilizing both dust and molecular hydrogen.* Note in SigChar W the ESO article that the star formation regions, "are characterized by comparatively high concentration of molecular gas and dust." *[24]* The question is how are the hydrogen molecules and dust utilized to facilitate star formation? It has been known for about 65 years that when high energy cosmic ray protons strike the Earth's atmosphere they each generate muons. The Stanford SLAC website reports that a count of muons arriving at the Earth's surface from one cosmic ray proton collision totaled over 1,000 muons. For the purposes of this paper, it is estimated that a $10^{15}$ eV proton collision in the Earth's atmosphere would generate hundreds to thousands of muons. It is well known that negative muons have the same negative charge as the electron, weigh about 207 times as much and can form muon-orbiting protons as atoms or ions. It is also known that a muon can orbit one proton forming an atom or it can orbit a proton-proton pair forming a molecular ion with the muon closely orbiting the proton pair because of its high mass, thereby pushing the protons closer together. It is also known that if one or both of the protons were replaced by deuterium, fusion would take place generating enormous amounts of energy, but the muon would be ejected unscathed to be able to catalyze additional fusion reactions. *[25][26]* With the mega flux of muons and electrons created by relativistic DM protons and helium nuclei bombarding the dust particles, helium and hydrogen, and the large flux of DM relativistic protons and helium nuclei passing through the region, a variety of particle collisions and particle configurations are possible in this high velocity muon-electron-proton plasma to trigger hydrogen fusion. For example, high-velocity high-collision-cross-section helium nuclei, being orbited by at least one muon, possibly could collide with a pair of protons being orbited by at least one muon, thereby triggering hydrogen fusion. The success of the fusion triggering action would depend upon how well the negative muons shield the positive charges of the nuclei involved so as to reduce the opposing coulomb forces sufficiently, thereby facilitating high collision velocities of the charged-shielded nuclei.

For additional information about cosmic rays and ultra-high-energy cosmic rays, see references *[18] [26] [27] [287] [29],* which include articles by A. A. Watson and by J. W. Cronin. For a published summary of the status of dark matter research as of early 2004 by Sir Martin Rees, see reference *[30]*.

**Tentative Conclusions, Insights, Explanations and Possible Astrophysical Discoveries**

A new research hypothesis has been developed by the author based upon finding astronomically based "cosmic constituents" of the Universe that may be created or influenced by or have a special relationship with possible dark matter candidates. A list of 14 relevant and plausible "cosmic constituents" of the Universe was developed by the author, which was then used to establish a number of constraints regarding the nature and characteristics of the long-sought dark matter particles. A dark matter candidate was then found that best conformed to the constraints established by the 14 "cosmic constituents."

The author will show in this section that this same conforming dark matter candidate provides plausible explanations for and discloses possible astrophysical discoveries related to the following astrophysical phenomena: *(1) the accelerating expansion of the Universe, (2) dark energy, (3) the "knees" and "ankle" of the cosmic ray energy spectrum graph, (4) the low star formation rates of LSB dwarf galaxies, (5) the ignition of hydrogen fusion reactions in the first generation stars, (6) the source of magnetic fields in spiral galaxies, (7) how dust particles facilitate hydrogen fusion in stars and (8) the four locations within the Local Group and the Virgo Supercluster from which earthbound cosmic ray protons depart.*

With the conforming dark matter candidate identified, it and its 24 Signature Characteristics will be utilized and applied, in the following paragraphs, to the 14 "cosmic constituents" to provide plausible explanations for the eight astrophysical phenomena highlighted in the previous paragraph. Each of the following 14 paragraphs begins with description of one of the 14 "cosmic constituents" shown in italics, followed by the alphabetic identification of the Signature Characteristics that are relevant to the conforming dark matter candidate.

*Relativistic Proton dark matter particles could:*

(1) *Form spherical DM halos around galaxies and DM halos around galaxy clusters.* See SigChar C, D, E, G, L and N. Dark Matter halos around galaxies and galaxy clusters have outer diameters and "hollow" core diameters determined by the galactic and extragalactic magnetic field magnitudes and the energy spectrum of the relativistic protons. The kinetic energies of the protons orbiting galaxy clusters are probably between one and two orders of magnitude higher than those orbiting galaxies, as determined by the Larmor Radius equation and the size of a galaxy cluster compared to the size of a galaxy. Also, the author believes that the outer diameter and core size of DM halos are not significantly affected by the amount of galaxy mass enclosed.

(2) *Cause the accelerating expansion of the Universe and possibly store dark energy.* See SigChar C, H, I, N, U and V. The accelerating expansion of the Universe may come about as a result of kinetic energy loss of the relativistic dark matter protons, orbiting galaxy clusters and galaxies, from synchrotron radiation losses and from collisions with photons, dust, hydrogen and helium leading to a reduced relativistic mass of the receding galaxy clusters, causing them to speed up to maintain their momentum under the Law of Conservation of Linear Momentum. (Also, it is possible that the total kinetic energy and/or the energy equivalent of the relativistic mass of all the DM relativistic protons could be linked to the mysterious "dark energy" since a portion that same energy seems to be consumed in conjunction with the accelerating expansion of the Universe.)

(3) *Be transformed into low velocity hydrogen, protons or proton cosmic rays.* See SigChar O. Low velocity hydrogen and protons and relativistic cosmic ray protons are all prevalent in the Universe. The most logical source of all three of these types of ordinary matter would be relativistic dark matter protons that had lost kinetic energy through synchrotron radiation or collisions with dust, photons or hydrogen.

(4) *Create the magnetic fields within and around spiral galaxies.* See SigChar F and U. The creation of magnetic fields surrounding spiral galaxies requires the flow of electric charges through space. Relativistic dark matter protons orbiting galaxies will create such magnetic fields through the astrophysical dynamo effect.

(5) *Be concentrated in the long large curved filaments of dark matter (announced by NASA 9/8/04), which form galaxy clusters where the DM filaments intersect.* See SigChar P, W and X. Some relativistic dark matter protons are concentrated in curved long large dark matter filaments owing to the high relativistic velocities of the protons and to the weak extragalactic magnetic fields created by the astrophysical dynamo effect. The author believes that the DM filaments may be slightly curved portions of supercluster halos of DM protons, the widths of which are confined electrostatically by the presence, within the filaments, of proton-produced muons and electrons (muons decay into electrons, etc.). Further, the crashing of intersecting DM filaments could lead to debris of relativistic protons at various energies and electrons from muon decay and slower moving hydrogen, helium and protons—all the necessary ingredients to form galaxy clusters, galaxies and stars.

(6) *Create large mature spiral galaxies less than 2.5 billion years after the Big Bang.* See SigChar Q. Relativistic protons could create large mature spiral galaxies less than 2.5 billion years following the Big Bang if their galaxy clusters are created via the top-down process described in item (5) above.

(7) *Create spherical DM halos having predictable outer and "hollow" core diameters.* See SigChar A, B, C, D, E, F and G. Orbiting relativistic protons could create spherical DM halos having predictable outer-and "hollow"-core diameters determined by the kinetic energies of the relativistic protons and the galactic and extragalactic magnetic field strengths, through use of the Larmor Radius equation.

(8) *Provide angular momentum to spiral galaxies and their DM halos.* See SigChar C and R. Orbiting relativistic protons creating magnetic fields via the astrophysical dynamo effect eventually will achieve a steady state dynamical configuration with significant angular momentum, which can be transferred to a spiral galaxy and its DM halo under the Law of Conservation of Angular Momentum. Spiral galaxies embedded within or overlapping the "hollow" cores of DM halos (essentially orbiting proton streams) represent one such steady state configuration with a high angular momentum.

(9) *Create galaxies without a central DM density cusp.* See SigChar B, C, E and S. Orbiting DM relativistic protons under the influence of a magnetic field generally will be constrained in orbits and will not move toward the nucleus of a galaxy unless they have lost a large percentage of their kinetic energy, in which case they probably would collide with one of the multitudinous star systems they would encounter on their way toward the galaxy nucleus. Therefore, the protons would not normally form a dark matter mass density cusp at the nucleus of a galaxy.

(10) *Create a starless galaxy or an LSB dwarf galaxy with low star formation rates.* See SigChar A, B, C, D, E, G, J, L, M and T. The author believes that the diameter of a galaxy embedded within the "hollow" core of a DM halo can be larger, smaller or the same size as the inner diameter of the "hollow" core. It is known that the galactic magnetic field strength of spiral galaxies is higher than the extragalactic magnetic field surrounding the DM halo. The low star formation rates for starless galaxies and LSB dwarf galaxies can be explained if the orbiting relativistic protons in the DM halo are normally utilized either to ignite the hydrogen fusion reaction of new stars or to provide proton fuel to the stars. It would follow that the low star formation rate could occur if the galaxy diameter is smaller than the halo core diameter and, therefore, protons in the DM halo would not be subjected to the higher magnetic field of the galaxy and to the subsequent synchrotron radiation losses which would normally cause them to move into the galaxy and thereby facilitate the formation of stars. Also, LSB dwarf galaxies are known to be isolated in space, which would eliminate dust accreted from external sources, which would lead to a lower star production rate. A dark galaxy can also exhibit a low star production rate if the DM halo magnetic field is lower than normal and/or the energies of the DM halo protons are higher than normal since in both these cases the DM halo protons would tend to remain in the DM halo. (Starburst galaxies, which have high star creation rates, represent the opposite case. They contain large amounts of dust with which the DM halo protons could collide, lose kinetic energy and move into the galaxy as their synchrotron radiation losses rise, thereby providing high energy proton fuel for star creation. Also, the dust leads to the production of muons, which is a known catalyst of hydrogen fusion.)

(11) *Lead to linearly rising rotation curves for LSB dwarf galaxies and to flat rotation curves for spiral galaxies.* See SigChar L. The recently announced (February 11, 2005 *[17]*) linearly rising rotation curves of LSB dwarf galaxies, compared to the flat rotation curves for large spiral galaxies, indicated to the researchers that the dark matter of LSB dwarf galaxy halos is "weakly centrally concentrated." This supports Drexler's previously developed concept of the "hollow" cores of DM halos and the core-size relationship to the size of the enclosed galaxies. One can thus conclude that if the enclosed galaxy is smaller than the "hollow" core diameter of the halo, the star formation rate probably will be low. If the galaxy disc diameter overlaps the "hollow" core of the DM halo, the galaxy probably will have a high star formation rate. If a galaxy is smaller than a typical LSB galaxy or if its galactic magnetic field is unusually low, it could appear to be a "dark galaxy." A dark dwarf galaxy can remain a "dark galaxy" accreting protons and helium nuclei from its DM halo for billions of years until it grows sufficiently in size to overlap the "hollow" core of its DM halo and star production could begin. A high level of dust particles in a galaxy and halo would probably bias it toward a high rate of star production because more protons from the DM halo would enter the galaxy and their collisions with dust would create muons, which catalyze hydrogen fusion.

(12) *Form about 80% to 90% of the mass of the Universe, the remainder being hydrogen, helium, etc.* See SigChar A, O and V. What dark matter particles could form about 80% to 90% of the mass of the Universe, the remainder being hydrogen and helium molecules, atoms and ions? The relativistic proton fits these astronomical data constraints since it can play a significant role both with the 80% to 90% dark matter and with the hydrogen and helium categories. It fits the hydrogen category easily since relativistic protons decay naturally into hydrogen. It also has a high relativistic mass such that even if it comprises only one tenth of the number of *particles* represented by the hydrogen, on a mass basis it still could comprise about 80% to 90% of the *mass* of the Universe. That is, it would meet this condition if its average particle mass were about 60 times the rest mass of a proton, represented by an average proton kinetic energy of only $5.5 \times 10^{10}$ eV. As is well known, slow-moving baryons have been ruled out as a dark matter candidate for many years because the primordial nucleosynthesis calculations indicate a low *particle abundance* of baryons in the Universe. However, this argument doesn't address the large quantity of relativistic protons in the Universe. This quantity of relativistic protons satisfies the *particle abundance* constraint and at the same time it also satisfies the *mass abundance* constraint as represented by the fact that dark matter particles, having a large relativistic mass, could enable them to form 80% to 90% of the mass of the Universe.

(13) *Ignite hydrogen fusion reactions of the first generation stars using only hydrogen atoms and of second generation stars utilizing hydrogen molecules and dust as well.* See SigChar A, K, W and X. *(a) Star ignition with only hydrogen atoms:* It is known that dwarf galaxies having a large amount of hydrogen and dark matter do not necessarily have a high star formation rate. Also, many astrophysicists believe that "to explain primordial star formation without molecular [hydrogen] gas and dust is a major challenge in modern astrophysics." See SigChar W and *[24]*. Yet astrophysicists know that primordial star formation did take place. The author believes that the relativistic proton dark matter theory offers a plausible explanation. The theory posits that the high velocity collisions by, for example, $10^{15}$ eV relativistic DM protons (and helium nuclei) with hydrogen particles in compressed hydrogen clouds also would generate muons to act as catalysts to facilitate a hydrogen fusion reaction and the ignition of new stars. That is, the muons could provide a partial coulomb shield for the $10^{15}$ eV protons (and helium nuclei) involved in collisions with protons so as to sufficiently increase their collision velocities to trigger hydrogen fusion. Note that this proton energy is at least 1,000 times greater than that produced by man-made accelerators. *(b) Star ignition with hydrogen gas molecules and dust would generate even more muons:* The relativistic proton dark matter and accompanying helium nuclei would enter into collisions with the hydrogen gas and dust particles. The author believes that collisions would generate muons which could replace some of the electrons in the hydrogen atoms, hydrogen molecules and in hydrogen/helium molecules, thereby forming ions with a muon orbiting two protons or a proton-helium pair, pushing them closer together and creating a collision target. The relativistic protons and accompanying helium nuclei could also capture muons, which would partly shield their coulomb charges and thereby increase their collision velocities to facilitate hydrogen fusion and star ignition.

(14) *Create the first "knee" at $3 \times 10^{15}$ eV, the second "knee" between $10^{17}$ eV and $10^{18}$ eV and the ankle at $3 \times 10^{18}$ eV of the cosmic-ray energy spectrum near the Earth*. See SigChar C, D and M. The author's relativistic proton dark matter theory posits that the highest energy DM protons are orbiting galaxy clusters and superclusters, a somewhat lower energy relativistic proton group orbits galaxies and a lowest energy relativistic proton group circulates within the enclosed galaxies. From the proton-flux versus proton-energy spectrum measured near the Earth and the Larmor Radius equation, one can determine that the proton flux density is highest within the Milky Way, next highest within the Milky Way's DM halo and lower in the DM halo of the Local Group of galaxies and lowest in DM halos associated with the Virgo Supercluster. The author believes that the first "knee" of the energy spectrum graph at $3 \times 10^{15}$ eV probably comes about because the protons that arrive at the Earth from the Galaxy have energies below the first "knee" energy level and those that come from the Galaxy's DM halo have energies above the first "knee" energy level. The second "knee" at about $3 \times 10^{17}$ eV probably separates protons arriving from the Galaxy's DM halo from those in the DM halo of the Local Group, which could have energies at or above $3 \times 10^{17}$ eV. The ankle energy is at or above $3 \times 10^{18}$ eV. The author believes that the four regions of the energy spectrum graph have four different slopes because they represent relativistic protons arriving from four different locations─the Milky Way, its DM halo and the DM halo of the Local Group of galaxies and the DM halos around and within the Virgo Supercluster. For example, since the Milky Way is about 60 million light-years from the center of Virgo it would not be surprising, under the author's theory, that only 3 to 4 protons per square kilometer per century would arrive at the Earth with energies above $10^{19}$ eV. (Note that the diameter of the Milky Way is about 100,000 light-years, its DM halo has an outer diameter of about 1 million light-years and its Local Group of galaxies has an outer diameter of about 10 million light-years and the diameter of the Virgo Supercluster is about 120 million light-years. Also note that the diameters of each of these four celestial bodies is about one order of magnitude greater than the one it encloses, which provides clues as to the kinetic energy levels of the orbiting protons in each case and how these energy levels relate to the energy levels of the "knees" and "ankle.")

**About This Paper**

Considerable effort was made not to bias the outcome of the Ockham's Razor logical search for a dark matter candidate fitting the intrinsic DM description laid down by the "cosmic constituents" and their related constraints. No dark-matter-linked "cosmic constituent" was knowingly excluded. And the author also attempted to maximize the number of "cosmic constituents" utilized, to assure more confidence in the DM selection process.

The cold dark matter theory has provided the only accepted theoretical dark matter candidates─WIMPs and neutralinos─during the past 20 years, but unfortunately neither of these particles has been detected anywhere in the Universe using either astronomical or particle accelerator techniques. After 20 years, it is necessary that other dark matter candidates be given an equal opportunity to compete using equal standards of acceptance. Further, the following astrophysical phenomena do not seem to be explainable by the cold dark matter theory of WIMPs and neutralinos: *(1) the accelerating expansion of the Universe, (2) dark energy, (3) the "knees" and "ankle" of the cosmic ray energy spectrum graph, (4) the low star formation rates of LSB dwarf galaxies, (5) the ignition of hydrogen fusion reactions in the first generation stars, (6) the source of magnetic fields in spiral galaxies, (7) how dust particles facilitate hydrogen fusion in stars and (8) the four locations within the Local Group and the Virgo Supercluster from which earthbound cosmic ray protons depart.*

The CDM theory's proponents provide no evidence as to the identification of or the nature of the CDM particle or its origin. Its acceptance relies on computer simulations. Do the computer simulation solutions represent the large-scale structure of the Universe formed by 100 billion galaxies or by 100 billion WIMPs? The theory and research rely entirely on gravitational forces and do not consider the charged particles (and the very high electromagnetic forces acting on them) that create the galactic and extragalactic magnetic fields. There is no CDM explanation for NASA's announcement on September 8, 2004 that in addition to forming halos around galaxies and galaxy clusters, dark matter is concentrated in long large DM filaments which form galaxy clusters where the filaments intersect. NASA's report also indicates a top-down theory of galaxy cluster formation while the CDM theory promotes a bottom-up theory of galaxy cluster formation. With dark matter representing about 80% to 90% of the mass of the Universe and dark energy representing 73% of the energy/mass of the Universe, it is difficult to imagine there is no link between the two, yet CDM theory does not address that possibility.

Also, the CDM theory does not seem to explain any of the eight astrophysical phenomena shown above in italics. This paper was written and the relativistic proton dark matter theory was developed to address these voids in the CDM theory while at same time recognizing CDM theory's success in explaining the formation and dynamics of large-scale structure of the Universe. (It is possible that the success of the gravitation-based CDM theory with regard to large-scale structures is based upon the fact that the effects of charged-particle forces may involve primarily short range particle dynamics and, therefore, could average out over a galaxy and over a galaxy cluster and also that input data to and solutions derived from computer simulations are scale independent.)

The selection of the relativistic proton as a promising dark matter candidate was based upon astronomical data (the 14 "cosmic constituents"), the laws of physics, Ockham's Razor logic and the concept of constraints on the nature and characteristics of the DM candidates. The selection process disregarded any theories regarding the nature of the early Universe and its particle zoo that have not been confirmed astronomically, because *unproven theories are in essence assumptions* and Ockham's logic favors those explanations which require the fewest assumptions.

I wrote a small book that was published as a paperback and e-book in December 2003 entitled, "How Dark Matter Created Dark Energy and the Sun," which presented an earlier version of this theory. From the critique of the book, I concluded that the earlier presentation of the theory relied on too many unproven theories having to do with the early Universe, which also created controversy. I decided to seek another approach relying entirely on modern astronomical data, which after many months led to this paper. It is hoped that the combination of Ockham's Razor logic, the laws of physics and the concept of constraints can be applied to other astronomical data to arrive at new astrophysical solutions.

## Acknowledgments

I thank George Blumenthal, Bruce Woodgate, Kim Griest, Alan Watson, Joel Primack, Sandra Faber and David Kliger for their critique of my theories or my book, "How Dark Matter Created Dark Energy and the Sun," which represents the 2003 version of my astrophysical theories.

I thank Sir Martin Rees for including my book amongst the five recommended for reading in the field of cosmology in conjunction with his TV science program in the United Kingdom, "What We Still Don't Know."

For contributing to my understanding of the nature of cosmic rays or the astrophysical dynamo effect, I acknowledge the writings of A. A. Watson, R. Clay, B. Dawson, J. W. Cronin, P. Sokolsky, S. Yoshida, G. Rudiger, R. Hollerbach, and A. M. Hillas.